\documentclass[11pt]{article}
\usepackage{epsfig}

\newcommand{\be}{\begin{equation}}
\newcommand{\ee}{\end{equation}}
\newcommand{\ba}{\begin{eqnarray}}
\newcommand{\ea}{\end{eqnarray}}
\newcommand{\baa}{\begin{array}}
\newcommand{\eaa}{\end{array}}
\newcommand{\nn}{\nonumber \\}

\def\mn{{\mu\nu}}

\def\tr{{\rm tr}}
\def\Tr{{\rm Tr\,}}

\def\CL{{\cal L}}                       

\def\tr{{\rm tr}\,}
\def\Tr{{\rm Tr}\,}

\newcommand{\nr}[1]{(\ref{#1})}   
\newcommand{\zt}{{\tilde z}}      
\newcommand{\rmi}[1]{{\mbox{\scriptsize #1}}}  

\begin{document}

\begin{titlepage}
\begin{flushright}
HIP-2006-42/TH\\
hep-ph/0609254\\
\end{flushright}
\begin{centering}
\vfill

{\Large{\bf Thermodynamics of AdS/QCD}}

\vspace{0.8cm}

K. Kajantie$^{\rm a}$         \footnote{keijo.kajantie@helsinki.fi},
T. Tahkokallio$^{\rm a,b}$      \footnote{touko.tahkokallio@helsinki.fi},
Jung-Tay Yee$^{\rm b}$         \footnote{jung-tay.yee@helsinki.fi}.

\vspace{0.8cm}

{\em $^{\rm a}$%
Department of Physics,
P.O.Box 64, FIN-00014 University of Helsinki,
Finland\\}

{\em $^{\rm b}$%
Helsinki Institute of Physics,
P.O.Box 64, FIN-00014 University of Helsinki,
Finland\\}

\vspace*{0.8cm}

\end{centering}

\noindent
We study finite temperature properties of  four dimensional QCD-like
gauge theories in the gauge theory/gravity duality picture. The
gravity dual contains two deformed 5d AdS metrics, with and without
a black hole, and a dilaton. We study the thermodynamics of the 4d
boundary theory and constrain the two metrics so that they
correspond to a high and a low temperature phase separated by a
first order phase transition. The equation of state has the standard
form for the pressure of a strongly coupled fluid modified by a
vacuum energy, a bag constant. We determine the parameters of the
deformation by using QCD results for $T_c$ and the hadron spectrum.
With these parameters, we show that the phase transition in the 4d
boundary theory and the 5d bulk Hawking-Page transition agree. We
probe the dynamics of the two phases by computing the
quark-antiquark free energy in them and confirm that the transition
corresponds to confinement-deconfinement transition.

\vfill
\noindent

%

\vspace*{1cm}

\noindent

\today

\vfill

\end{titlepage}

\section{Introduction}
In the gauge-gravity duality picture the gravity dual of finite
temperature ${\cal N}=4$ supersymmetric Yang-Mills theory is a
5-dimensional AdS space (times $S_5$) with a black hole
\cite{maldacena,witten}. While this is a very special conformally
invariant theory and requires also $g^2N_c\gg 1, N_c\gg1$ for its
validity, it has recently become apparent that one could
meaningfully apply the gauge-gravity duality picture to QCD matter
produced in relativistic heavy ion collisions, to viscosity
\cite{pss,kss,ks}, to jet energy loss
\cite{lrw,hetal,h,buchel,gubser,nakano} and to photon production
\cite{photon}. Note that these are all quasistatic phenomena; the
gravity dual of a full dynamic heavy ion collision is unknown, for
attempts, see \cite{nastase,ssz,jp}.

Ideally, one would like to derive the gravity dual of QCD in a
string theory framework, but this top-down way has so far not lead
to a unique result. Instead, one has, in a phenomenological
approach, studied various deformed metrics as candidate gravity
duals of QCD-like theories
\cite{brodsky,erlich,rold1,bigazzi,rold2,ghoroku,katz,karch}. The
purpose of this article is to continue this line and to present a
gravity dual model for QCD thermodynamics. The model will cover both
the deconfined phase (stable at $T>T_c$, metastable at $T<T_c$), the
confined phase (stable at $T<T_c$, metastable at $T>T_c$) and the
phase transition in between. The model for the high $T$ phase will
be a deformed AdS black hole metric with a dilaton, the model for
the low $T$ phase will be a deformed horizonless AdS metric with a
dilaton. A deformation is defined as a multiplication of the 5d
metric by an arbitrary function depending on the fifth coordinate
$z$. The dilaton plays an essential role here. The phase transition
will be analyzed in two independent ways: by computing the energy
momentum tensor in the 4d boundary theory and by computing the
difference of the 5d bulk actions
\cite{hawkingpage,witten,mmt,herzog2}. The shapes and parameters of
the deformations of the two phases are fixed by using the QCD number
for the $T_c$ of the 4d boundary theory and by using constraints
from the $T=0$ hadron spectrum \cite{karch}. With these parameters
the 5d bulk Hawking-Page transition coincides with the 4d boundary
one. With the same parameters one can then also compute the $Q\bar
Q$ free energy (see also \cite{az,braga} in the two phases. The
significant role of the dilaton is apparent here, the deformation in
the string frame is qualitatively different from that in the
Einstein frame. At $T=0$ we see that the $Q\bar Q$ potential
contains a confining linear part but also string breaking,
indicating that the deformation contains effects of dynamical
quarks. At $T>T_c$ the $Q\bar Q$ pair becomes deconfined with
entropy $\approx$ 2.1/pair.

The model we end up with is admittedly not unique and contains several
parameters, but this is
due to the fact that we wish to treat a large number of different
QCD phenomena at the same time. Limiting oneself to one phenomenon at
a time, simpler gravity duals can be presented. Our result shows that
a comprehensive solution is possible.

The paper is organized as follows. In section 2, we specify the
model. We assume a five dimensional gravity action
with a dilaton and a matter Lagrangian and postulate two
solutions corresponding to QCD-like theories on the boundary. In
section 3, using gravity/gauge theory duality we compute the energy
momentum tensor on the 4d boundary and find the phase structure of the
boundary gauge theory.
In section 4, we use the meson mass spectrum to constrain the
parameters of our approach.
In section 5, we consider a Hawking-Page type
transition of the 5d bulk geometry and find the correspondence between
the phase transition on the boundary and the phase transition in the
bulk. In section 6, we compute the
Wilson-Polyakov loops and correspondingly the free energy (potential
at $T=0$) between
a probe quark and antiquark. This confirms that, indeed, at high
temperatures we are in the deconfined phase and at low temperature in
the confined phase. Section 7 contains a discussion.

The numbers related to experimental ones quoted below should be
understood to have a sizable ${\cal O}$(10\%) error.
\section{The Model}

We consider the following five dimensional action in Einstein frame,
in standard notation:
 \be
  \label{gravityaction}
  S={1 \over 16 \pi G_5} \int d^5x \sqrt{-g} \left(R-2 \Lambda-{4\over3}
  (\partial_\mu \Phi)^2 \right)
  -{1\over 8 \pi G_5} \oint d^4x \sqrt{-\gamma}\Theta +\int d^5x \sqrt{-g}
  e^{a \Phi} L_m.
 \ee
The second term is the boundary term (for a boundary surface
$z=\epsilon$) and the third term is the matter part, which we do not
have to specify completely. The constant $a$ is dependent on the
coupling of matter to gravity. This action can be the bosonic part
of a supergravity action descendant from string theory. Note that
even though we do not specify the origin of the model in detail, we
have in mind near horizon geometry of stacked branes solution that
gives asymptotically $AdS_5$ space. A nontrivial matter part can be
introduced if we add, for example, a spacetime filling brane adding
fundamental matter to the
problem\cite{karchkatz,babington,mateos,sugimoto}. The effect of the
presence of fundamental matter would be already reflected in the
full supergravity solution.

We assume that $L_m$ can be so chosen that the following metric is a
solution as a dual to finite temperature QCD.
 \be
  ds^2 =  \CL^2 {e^{h(z)} \over z^2} \left[-g(z) dt^2
  + dx_1^2 +dx_2^2+dx_3^2+
   {dz^2 \over g(z)}   \right],
   \label{bulkmetric}
 \ee
with a dilaton $\Phi(z)$. The function $h(z^2),\, h(0)=0$ defines
the deformation. The boundary theory lives at $z=0$ and the geometry
would be asymptotically (when $z\to0$) AdS. We shall define the
string frame so that it contains the combination
$\sqrt{-g}e^{-2\Phi}R$ in the action. The transformation between
Einstein frame and string frame is then given by
 \be
   g_{MN}^{(E)}= e^{-{4\over 3} \Phi} g_{MN}^{(s)}
 \ee
and the bulk metric in the string frame will be the same as \nr{bulkmetric}
but with
\be
e^{h(z)}\Rightarrow e^{f(z)},\qquad f(z)\equiv h(z)+{4 \over 3}\Phi(z).
\label{fz}
\ee

\begin{figure}[!htb]
\begin{center}
\includegraphics[width=0.4\textwidth]{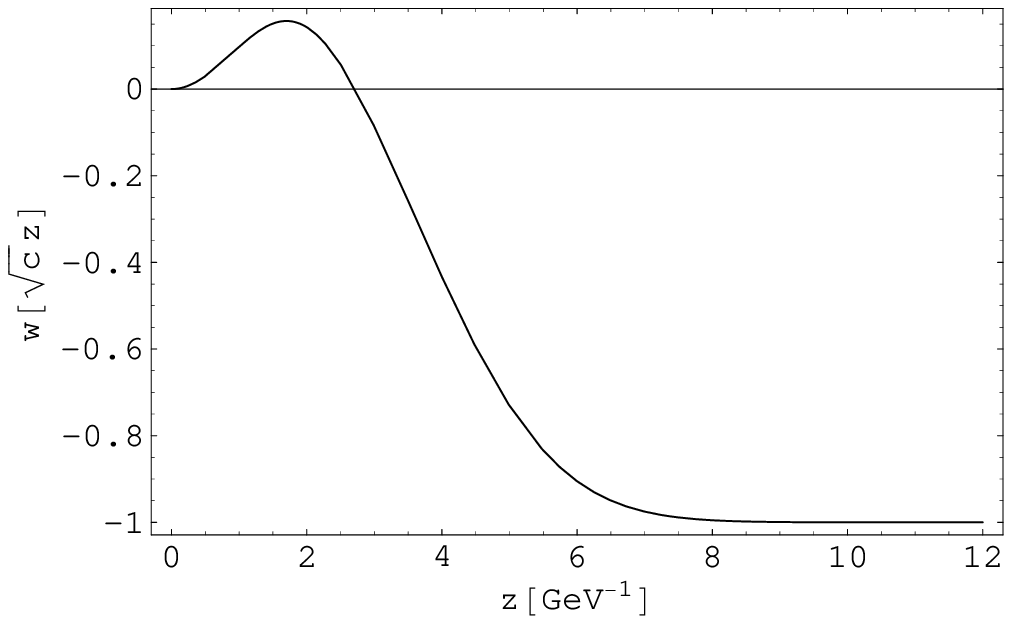} \hspace{1cm}
\includegraphics[width=0.4\textwidth]{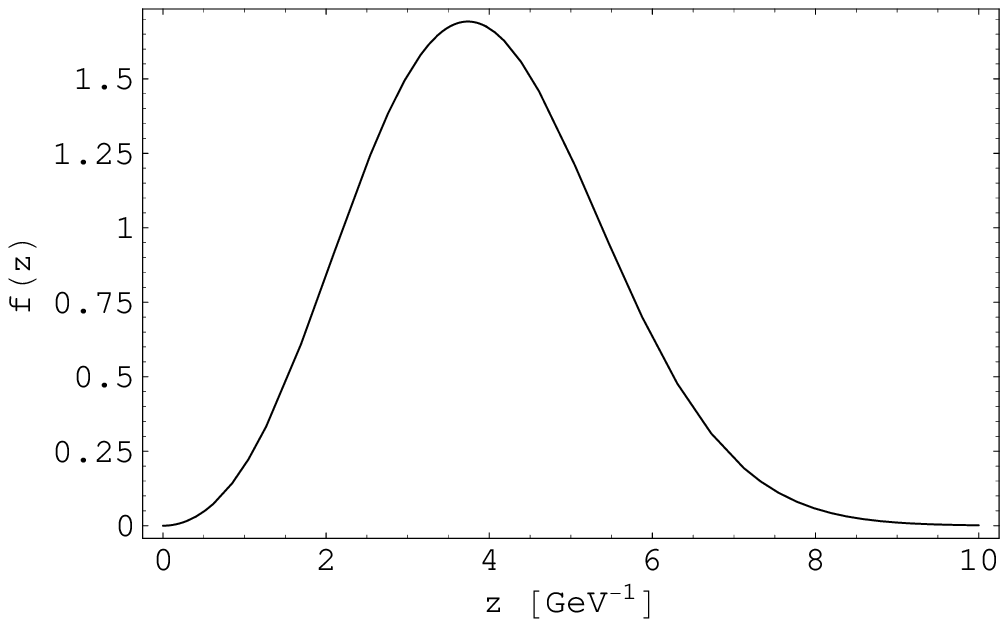}
\end{center}
\vspace{0 cm} \caption{\small A plot of $w(z)$ defined by
Eq.\nr{trialw} (left) and a plot of the low $T$ metric string frame
deformation $f(z)=29[w(\sqrt{c}z)-w(\infty)]cz^2/20$, $c=0.127$
GeV$^2$, $w(\infty)=-1$. }\label{wz}
\end{figure}

Our gravity dual is defined by specifying the deformation $h(z)$ and
the dilaton $\Phi(z)$. We will have two alternatives, one will be the stable
phase for $T>T_c$, the other for $T<T_c$. The former is
 \ba
 \label{hightempmetric}
  h(z) &=& -c z^2  \nn
  \Phi(z) &=& {3 \over 4} \phi z^2  \nn
  g(z) &=& 1-{z^4 \over z_0^4},
 \ea
and the latter
 \ba
 \label{lowtempmetric}
  h(z) &=& -c z^2 (1- {29\over 20} w(\sqrt{c}z))  \nn
  \Phi(z) &=& {3 \over 4} \phi z^2 \nn
  g(z) &=& 1.
 \ea
Here \be z_0={1\over \pi T} \ee is the horizon radius of the metric
\nr{bulkmetric}, $c \approx 0.127$ GeV$^2$ is a constant determined
from the $T_c$ of the QCD phase transition in the 4d boundary theory
and $\phi \approx 0.311$ GeV$^2$ is a constant which, together with
$w(\infty)\equiv w(x\to\infty)=-1.0$ is determined from the mass
spectrum of QCD. A further property of the function $w(x)$ is that
it for small $x$ should behave as $w(x)=x^2+{\cal O}(x^4)$, also
constrained by the QCD phase transition in the 4d boundary theory
(to make the bag constant vanish in the low temperature phase). We
can thus constrain the UV and IR limits of $w(x)$; the intermediate
behavior affects details of the $Q\bar Q$ potential. We will use the
ans\"atz
 \be
  w(x)=2x^2e^{-x^2}-\tanh x^2, \label{trialw}
 \ee
plotted in Fig.\ref{wz}. In the same figure we also show the
combination $f(z)=29[w(\sqrt{c}z)-w(\infty)]cz^2/20$, later
identified as the deformation of the low $T$ metric in the string
frame Eq.\nr{flz}. Further important constraints come from the fact
that we want the phase transition in the 4d boundary theory and the
Hawking-Page transition in the 5d bulk theory to agree. In
particular, this is possible only if the metric deformations
coincide for $z\to0$, i.e., the constants $c$ and $\phi$ coincide in
\nr{hightempmetric} and \nr{lowtempmetric}. The 5d bulk transition
further constrains $c>0,\,w(\infty)<20/29$. Also note in the ansatz
\nr{hightempmetric} and \nr{lowtempmetric}, we have assumed the
dilaton and matter from $L_m$ have the same leading order behavior.

\section{Thermodynamics of the Gauge Theory on the Boundary}\label{thermo}
We know that the metric (\ref{bulkmetric}) for $h=0$ is a solution
of the 5d AdS equations
 \be
  R_{MN}-{1\over2}Rg_{MN}-{6\over\CL^2}g_{MN}=0 \label{ads5}
 \ee
and the boundary theory is hot ${\cal N}=4$ supersymmetric
Yang-Mills theory in 4d at the temperature $T=1/\pi z_0$ with a
pressure = 3/4 times the ideal gas value\cite{gkp}. How is this
affected by the introduction of the deformation $h(z)$? To be able
to treat the two cases \nr{hightempmetric} and \nr{lowtempmetric}
simultaneously, write $h(z)=d_1 z^2+ d_2 z^4+ {\cal O}(z^6)$, only
the parameters $d_1,d_2$ of dimension $T^2,T^4$ matter for this
problem.

To answer this question we should compute the energy-momentum
tensor of the boundary theory. Since the theory is strongly coupled,
the computation directly from the boundary field theory is not feasible,
but we can use the gravity/gauge theory duality for this
purpose. How this is to be done has been studied, say,
in \cite{hs,balakraus,myers,skenderis,skenderis2}. However, the general methods
involve writing down an explicit 5d gravity-matter action and obtaining the metric and
other matter fields as solutions of equations of motion. Here our starting point is
a phenomenological deformed metric and the gravity action contains an unspecified
matter Lagrangian. This unknown part will lead to an unknown part in the energy
momentum tensor we obtain. It will cancel when studying the phase transition,
computing the difference of pressures
between two phases. Analoguously, in the later (Section 5) analysis of the
phase transition as a bulk Hawking-Page transition in 5d, the unknown matter Lagrangian
can be eliminated using equations of motion.

We transform the metric \nr{bulkmetric} to the form
 \be
   ds^2=\frac{\CL^2}{\zt^2}\big(g_{\mu\nu}(x,\zt)dx^\mu
   dx^\nu+d\zt^2\big),
   \label{fefferman}
 \ee
where $g_\mn\to \eta_\mn$  at $\zt=0$. If we expand the metric
$g_\mn$ for small $\zt$:
 \begin{equation}
  g_{\mu\nu}(\zt,x)=\eta_{\mu\nu}(x)+g^{(2)}_{\mu\nu}(x)\zt^2+
  g^{(4)}_{\mu\nu}(x)\zt^4+
  g^{(6)}_{\mu\nu}(x)\zt^6+\ldots,
 \end{equation}
the energy-momentum tensor $T_\mn$ of the boundary theory is given
by \cite{skenderis}
 \begin{equation}
  T_{\mu\nu}(x)=\frac{\CL^3}{4\pi G_5}[g^{(4)}_{\mu\nu}(x)+X_\mn+Y_\mn]
  ={N_c^2\over 2\pi^2}[g^{(4)}_{\mu\nu}(x)+X_\mn+Y_\mn], \label{skenderis}
 \end{equation}
where
 \ba
   X_\mn&=&-{1\over8}\eta_\mn[(\tr g_{(2)})^2-\tr
   g_{(2)}^2]-{1\over2}g_{(2)\mn}^2+ {1\over4}\tr g_{(2)}\cdot
   g_{(2)\mn}\nn
   &=&-C^2\eta_\mn\quad{\rm if}\,\,g_{(2)\mn}=C\eta_\mn
   \label{xmn}
 \ea
depends only on the square of contractions of $g_{(2)}$ and where
\be
Y_\mn=C_m\eta_\mn
\label{ymn}
\ee
is a part depending on matter fields.
We also used that for AdS$_5\times S^5$ spacetime
 \be
   G_{10}=G_5\pi^3\CL^5={\pi^4\CL^8\over2N_c^2}.
   \label{norm}
 \ee
For a thermal stationary and homogeneous system one, of course, has
$g_{\mu\nu}(\zt,x)=g_{\mu\nu}(\zt)$.

Although we cannot determine the constant $C_m$ in \nr{ymn}, we expect
it to be the same for both phases. This is analoguous to the fact that
in our metric ansatz \nr{hightempmetric} and \nr{lowtempmetric} the
leading boundary behaviour of both the deformation $h(z)=-cz^2+\cdots$ and
the dilaton $\Phi(z)={3\over4}\phi z^2$ is the same in both metrics;
which also imply that this holds for $g_{(2)\mn}$ computed below.

For the deformed metric \nr{bulkmetric} the transformation relating
$z$ and $\zt$ is defined by
 \be
   \int_\epsilon^\zt {d\zt\over\zt}=
   \int_\epsilon^z {e^{h(z^2)/2}\over\sqrt{g(z)}}{dz\over z},
 \ee
where we have enforced the condition $\zt\approx z$ for $z\to0$ by
putting the same $\epsilon$ as the lower limit in both integrations.
This can be integrated to
 \be
   \zt^2={2z^2\over 1+\sqrt{g(z)}}
  \exp\biggl[\int_0^z dz{e^{h(z)/2}-1\over
  z\sqrt{g(z)}}\biggr],
  \label{ztz}
 \ee
For $h=0$, unmodified black hole, one can invert
 \be
  z^2={\zt^2\over 1+\zt^4/(4z_0^4)}
 \ee
and get for the metric the exact result of \cite{jp}
 \be
  ds^2=  {\CL^2 \over \zt^2} \left[-{(1-\zt^4/4z_0^4)^2\over 1+\zt^4/4z_0^4} dt^2
  + (1+{\zt^4\over 4z_0^4})(dx_1^2 +dx_2^2+dx_3^2)+
   d\zt^2  \right].
   \label{jpmetric}
 \ee
For any $h$, to invert \nr{ztz} to express $z^2$ in terms of $\zt^2$,
we expand in $z^2$ using the expanded form
$h(z^2)=d_1z^2+d_2z^4+...$, higher terms do not matter. One then
obtains, for the high temperature metric \nr{hightempmetric} with the
horizon $z_0$
 \be
   g_\mn=\eta_\mn-{3\over2}d_1\eta_\mn \zt^2+[(u_\mu
   u_\nu+{1\over4}\eta_\mn){1\over z_0^4}-
   {20d_2-7d_1^2\over16}\eta_\mn]\zt^4+..,
 \label{gmunu}
 \ee
where $u^\mu=(1,0,0,0)$ so that $u_\mu u_\nu+{1\over4}\eta_\mn={\rm
diag}(3/4,1/4,1/4,1/4)$. Using \nr{skenderis},\nr{xmn},\nr{ymn} and
\nr{norm} the energy-momentum tensor then is, for the high
temperature metric,
 \be
 \label{boundaryemtensor}
  T_\mn=\left( \begin{array}{cccc}
  3aT^4+B_h & 0 & 0&0 \\
  0 & aT^4-B_h & 0 &0\\
  0 & 0 & aT^4-B_h &0\\
  0 & 0 & 0 & aT^4-B_h
  \end{array} \right),
 \ee
where
 \be
  a={\pi^2N_c^2\over8}\qquad
  B_h={N_c^2\over32\pi^2}(-20d_2+29d_1^2-16C_m)={N_c^2\over32\pi^2}(29c^2-16C_m),
  \label{bh}
 \ee
where we have inserted for the deformation of the high temperature metric
\nr{hightempmetric} $d_1=c,\,d_2=0$. For the low temperature metric
\nr{lowtempmetric} there is no $T^4$ term,
\be
T_\mn=-B_l\eta_\mn
\ee
with
\be
B_l={N_c^2\over32\pi^2}(-20d_2+29d_1^2-16C_m)={N_c^2\over32\pi^2}(-16C_m),
\label{bl}
\ee
where we have constrained the low temperature metric deformation by
$20d_2=29d_1^2$. The motivation for the omission of the $T^4$ term is that
in the QCD phase transition there is a large
change in the number of degrees of freedom when one crosses $T_c$.
In fact, at large $N_c$, $p_h'(T_c)\sim {\cal O}(N_c^2)$ while
$p_l'(T_c)\sim {\cal O}(1)$ \cite{pisarski}.
In the high $T$ phase there are gluons and quarks while the low $T$ phase
only contains massless pions. The motivation for the constraint
$20d_2=29d_1^2$ is that it minimises the vacuum energy in the
low temperature phase in which the hadrons move in the true physical
vacuum. For example, in the bag model for
hot QCD matter one writes
$p_h(T)=a_hT^4-B,\,\,p_l(T)=a_lT^4$ with $a_h\gg a_l$
\footnote{Though the lattice data rather follows the pattern
$p=a_hT^4-{\tilde B}T^2$ \cite{boyd}}.

\begin{figure}[!htb]
\begin{center}
\includegraphics[width=0.4\textwidth]{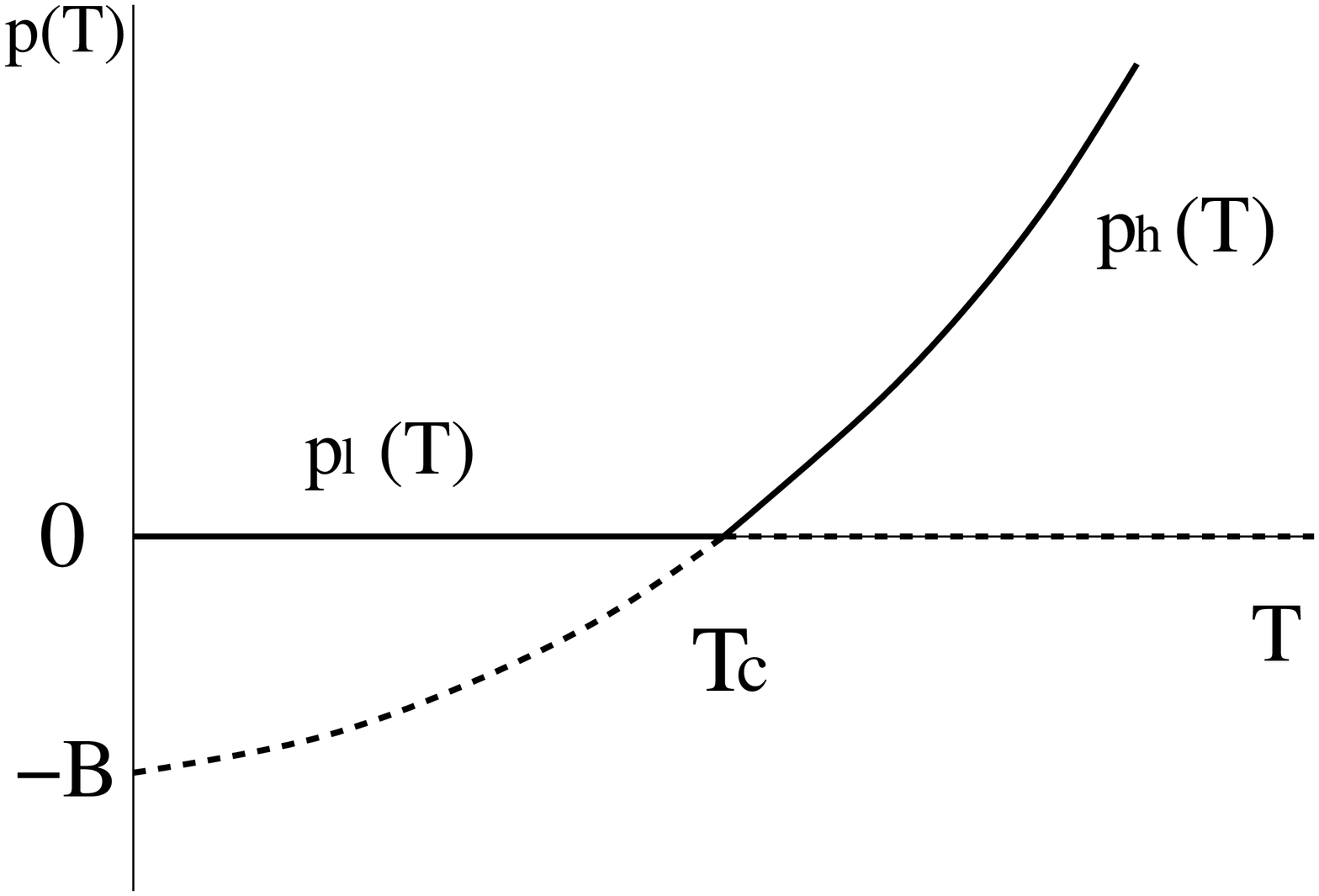}
\hspace{2 cm}
\includegraphics[width=0.4\textwidth]{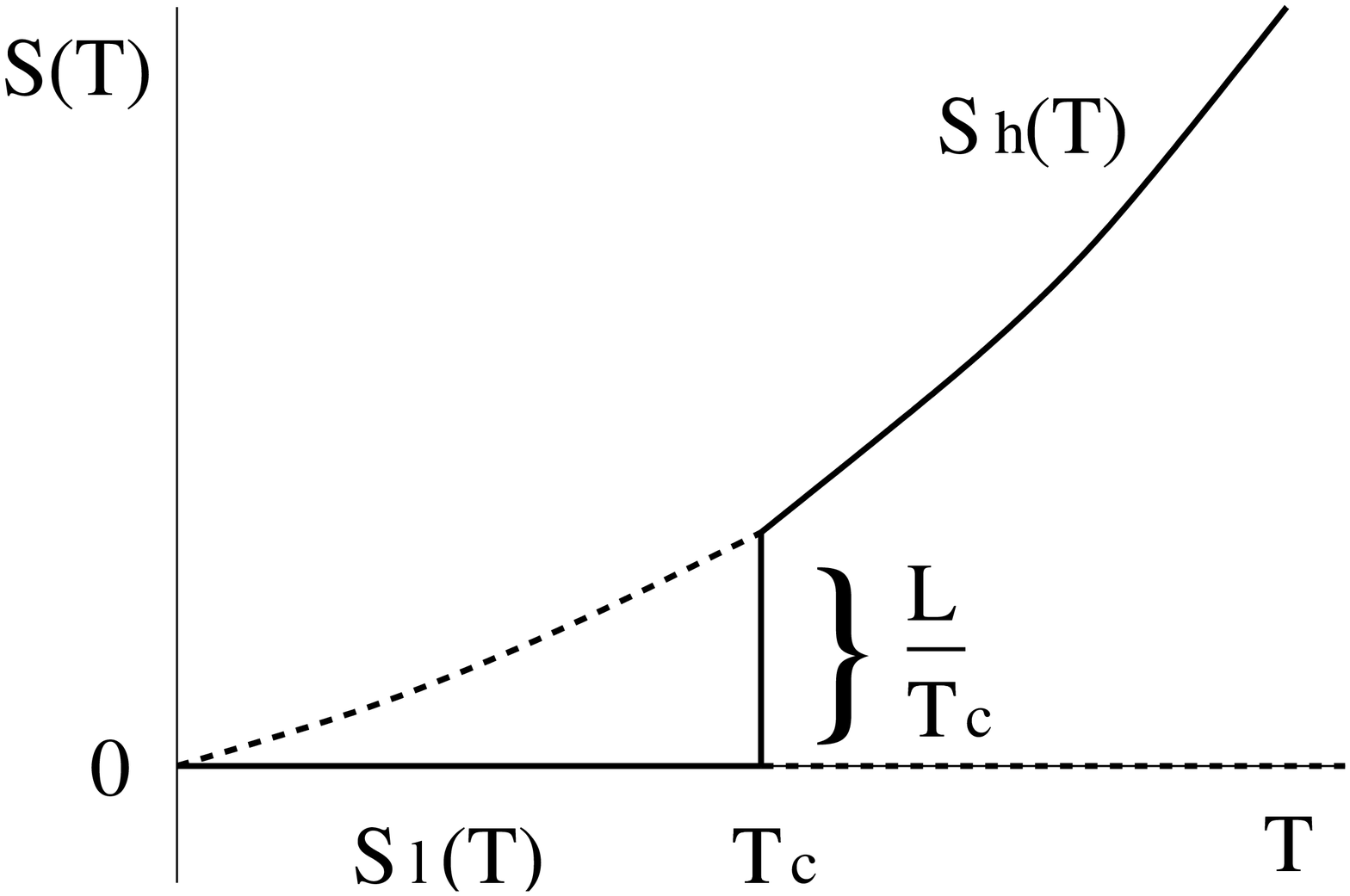}
\end{center}
\vspace{0 cm} \caption{\small The pressure $p(T)$ and entropy
density $s(T)=p'(T)$ in terms of temperature $T$ with $B=29
c^2N_c^2/(32\pi^2)$ and $L=T_c \Delta s=29 N_c^2 c^2/( 8 \pi^2)$.
The contribution of the matter fields is neglected here,
$C_m\ll c^2$. }\label{phasediagram}
\end{figure}

Thus our result for the pressures of the two phases
is
 \ba
 \label{finalpressure}
  p_h(T)&=&
    {\pi^2N_c^2\over8} T^4 -{N_c^2\over32\pi^2}(29 c^2-16C_m)\nonumber \\
  p_l(T)&=& {N_c^2\over32\pi^2}(16C_m),
 \ea
and for the entropy density
 \ba
 \label{finalentropydensity}
  s_h(T)&=&
    {\pi^2N_c^2\over2} T^3 \nonumber\\
  s_l(T)&=& 0,
 \ea

The phase transition occurs at $p_l(T_c)=p_h(T_c)$.
The unknown matter field contribution $C_m$ cancels from here, leading to
 \be
 \label{boundarycriticaltemp}
  \pi T_c=\left({29\over 4}\right)^{1/4} \sqrt{c}.
 \ee
This is what one dimensionally expects, introducing a deformation
$\exp(c z^2+ \cdots)$ one has, in addition to $T=1/\pi z_0$,
introduced a new energy scale $c^{1/2}$ to the problem. Using
\nr{boundarycriticaltemp} to determine the constant $c$ we have,
e.g.,
 \be
  T_c=186\,{\rm MeV}\Rightarrow c\approx 0.127\,{\rm GeV}^2.
 \ee

The general behavior of the pressure $p(T)$ and entropy density
$s(T)=p'(T)$ is shown in Fig. \ref{phasediagram}. The transition
corresponds to first order phase transition with two metastable
branches persisting at all $T$. In  section 6 we will show that the
stable branch $T>T_c$ corresponds to a deconfined phase and that at
$T<T_c$ to a confined phase.

For high temperature geometry (\ref{hightempmetric}), one can get
the entropy directly from the geometrical data of the bulk,
computing the area of black hole or evaluating the Euclidean action.
If we calculate the entropy
density from the area law, we get
 \be
  s_h(T) ={\pi^2N_c^2\over2} T^3 e^{-{3\over 2 \pi^2} {c\over T^2}}.
  \label{bhentropy}
 \ee
This looks very interesting phenomenologically since it fits well
with the lattice data\cite{boyd} with proper adjustment of number of
degrees of freedom. However, it differs from the result
(\ref{finalentropydensity}) for the entropy density. This is again
due to the fact that we start from a deformed metric and do not
derive it as a solution of 5d gravity-matter equations. For this
complete treatment the results should concide \cite{ps} changing
also the black hole entropy formula \nr{bhentropy}.

\section{Meson Mass Spectrum}

In this section, we show how it is possible within our framework
to get a mass spectrum for mesons which grows linearly with the
excitation level $n$ and spin $S$, i.e., $m^2\sim (n+S)$. We use
the constraints, considered first in \cite{karch}: in the limit
$z\to\infty$, the function $f(z)$ and the dilaton $\Phi(z)$ should
behave as $f(z)\to z^k$, where $k<2$, and $\Phi(z)\to z^2$, to
obtain a linear mass spectrum as a function of $n$ and $S$.

At zero temperature, we have the metric \nr{lowtempmetric}. We
specify the matter part in \nr{gravityaction} to contain a spin
$S$ field :
 \be
  S=\int d^5 x \sqrt{-g} e^{- \Phi} \left({1\over 2}(DA_S)^2+{1\over 2}M^2
  A_S^2\right),
 \ee
where $A$ is totally symmetric rank $S$ tensor field. We are
assuming that the matter in this case is open string matter on the
spacetime filling  branes, for example,  and the coupling to the
dilaton is fixed in this spirit. In the axial gauge, the equation of
motion for the transverse traceless part, after rescaling
$A_S=e^{(S-1)F(z)} \tilde{A}_S$ with $F(z)=f(z)-2\textrm{log}(z)$,
is
 \be
  \partial_z\left(e^{( S-{1\over 2})F} e^{- \Phi} \partial_z \tilde{A}_S
\right)+m^2 e^{(S-{1\over 2})F} e^{- \Phi}
  \tilde{A}_S=0.
 \ee
After the substitution $\tilde{A}_S = e^{B\over 2} \psi$ with
$B(z)= \Phi(z)- (S-{1\over 2})F(z)$, the equation is reduced to
the Schr\"{o}dinger form
 \be
  \left(-\partial_z^2+ V(z)\right) \psi = m^2 \psi
 \ee
with the potential
 \be
  V(z)={1\over 4} (\partial_z B)^2- {1\over 2} \partial_z^2 B.
 \ee
To have the expected behavior of mass $m^2 \sim n$, the asymptotic
form of the potential should be quadratic \cite{karch}. This
constrains the form of the metric strictly. We can read off of the
asymptotical mass spectrum of spin $S$ meson from
(\ref{bulkmetric},\ref{lowtempmetric}),
 \be
 \label{mesonmass}
   m^2=4\left[\frac{3}{4}\phi-(S-\frac{1}{2})\bigg(\phi
   - c (1-{29 \over 20} w(\infty))\bigg)\right](n+S),
 \ee
where $w(\infty)$ is the limiting value of the function $w(z)$ in
\nr{lowtempmetric} when $z\to\infty$. We have also used
$\Phi(z)=\frac{3}{4}\phi z^2$. To match the experimental data
\cite{eidelman}, the term multiplying $S-1/2$ must be canceled. This
fixes the value of the dilaton:
 \be
  \phi=c  \left(1-{29 \over 20} w(\infty)\right).
  \label{phi}
 \ee
For the $n$-excitations of the rho meson with spin $S=1$,
$I^GJ^{PC}=1^+1^{--}$ one obtains after this cancellation
\footnote{The near boundary structure of the function $w(z)$ affects
the exact mass spectrum \cite{karch}, and therefore needs a closer
analysis.}
 \be
  m^2_{\rho}\simeq 3\phi \,n.\label{mrho}
  \ee
A linear fit to the observations is $m^2_\rho=0.93 n $ GeV$^2$
\cite{eidelman}. Eq.\nr{mrho} then gives $\phi=0.31$ GeV$^2$ and,
using the earlier established value $c=0.127$ GeV$^2$, \nr{phi}
gives $w(\infty)=-1.0$.

It is natural to expect that $g_{YM}^2\sim e^{\frac{3}{4}\phi z^2}$.
In our case $\phi > 0$, and the coupling flows toward strong
coupling regime from UV to IR. Even though the direct comparison of
the weakly coupled and strongly couple regime is impossible, this is
still qualitatively consistent. In our notation, this also gives the
condition that $w(\infty) < {20\over 29}$.

\section{Hawking-Page Transition of the 5d Bulk
Geometry}

In section \ref{thermo}, we have computed the boundary energy
momentum tensor for the deformed geometry which corresponds to finite
temperature QCD on the boundary in terms of gravity/gauge theory
duality. Notice again that the energy momentum tensor
(\ref{boundaryemtensor}) is that of the 4d boundary theory and the
phase transition is based on the 4d boundary theory picture. In this
section we will try to compute the phase transition in the 5d bulk
\`{a} la Hawking and Page \cite{hawkingpage}\footnote{A similar idea
was used for the computation of deconfinement temperature in
\cite{herzog2} for a deformed AdS geometry.}. First we Euclideanize
the metric (\ref{bulkmetric}) performing the Wick rotation $t
\rightarrow i \tau$, then evaluate the Euclidean
action \nr{gravityaction} corresponding to the two metrics
\nr{hightempmetric} and \nr{lowtempmetric}. The metric corresponding
to the smaller action is the stable one.

The action \nr{gravityaction} contains the unspecified matter term $L_m$,
but actually this can be eliminated using the equation of motion of the dilaton.
The contribution to the action from the matter part is given by
\be
  S^E_m= -{1 \over 6a\pi G_5}\int d^5x\partial_z[\sqrt{g} g^{zz}
\partial_z \Phi]
=-{1 \over 6a\pi G_5} \int d\tau d^3x
\vert_{z=\epsilon}^{z=z_0,\infty}[\sqrt{g} g^{zz}
\partial_z \Phi],
 \ee
The upper limits $z=z_0$ and $z=\infty$ correspond to the metrics
(\ref{hightempmetric}) and (\ref{lowtempmetric}) and both give a vanishing
contribution. The contribution from $z=\epsilon$ is divergent but the
divergence is the same for the two metrics and thus cancels when the
difference between the actions is evaluated. Thus the $L_m$-term can be
entirely omitted from the discussion.

To evaluate the action for the remaining terms we have to specify the
limits of integration in $\int d^5x=\int d\tau dz d^3x$. The integration over
${\bf x}$ is over all the space and thus gives only the volume factor $V_3$.
The integration over $z$ is over $(\epsilon,z_0)$ for (\ref{hightempmetric})
and over $(\epsilon,\infty)$ for (\ref{lowtempmetric}). Since
we are assuming that the black hole is in thermal equilibrium with
thermal radiation, the high temperature geometry
\nr{hightempmetric} has a natural temperature defined by the black
hole temperature $\beta_h= \pi z_0$ and the integration over $\tau$ is
over the interval $(0,\beta_h)$. The temperature of thermal
radiation in \nr{lowtempmetric} can be arbitrary. For
comparison with the high temperature metric, the temperature
is fixed \cite{hawkingpage,witten} requiring
that both  have the same physical circumference along the Euclidean
time direction $\tau$ at the regularization point $z=\epsilon$.
This leads to the relation
 \be
  \label{temperature}
   \beta_l=\beta_h\sqrt{g_h/g_l}e^{(h_h-h_l)/2}\big|_{z=\varepsilon}=
   \beta_h \sqrt{1-{\varepsilon^4\over z_0^4}}
   e^{-{29\over 40} c \varepsilon^2 w(c\varepsilon^2)}\simeq \beta_h\bigg
[1-\bigg(\frac{1}{2z_0^4}+\frac{29c^2}{40}\bigg)\epsilon^4 + \mathcal{O}
(\epsilon^6)\bigg].
 \ee

Since the $\tau$-integrals just give the upper limits, the action difference
to be evaluated now is
 \ba
 \label{actiondifference}
  \Delta S^E &=& S^E_h - S^E_l= -{V_3\beta_h \over 16 \pi G_5}\left(
   \int_\varepsilon^{z_0} dz \, L_h
  -{\beta_l\over\beta_h}\int_\varepsilon^\infty dz \, L_l
  \right)+ \Delta S_{\rm boundary},
 \ea
where $L_h$ and $L_l$ are the  Lagrangian densities without the $L_m$-term
in \nr{gravityaction} evaluated with
the metrics \nr{hightempmetric} and \nr{lowtempmetric}, respectively.
The $z$-integrals in \nr{actiondifference} converge since $c>0$ and since the
earlier result $w(\infty)=-1.0$ satisfies $w(\infty) <20/ 29$.
The $\varepsilon\to0$ divergences in \nr{actiondifference} require a more
careful analysis. The divergent parts of both of the
Lagrangians are the same\footnote{This can be guaranteed by choosing
the same value for the parameters $\phi$ and $c$, in both metrics.}:
 \be
  L_{div}=-\frac{8}{z^5}-\frac{16 c}{z^3}+(27c^2-3\phi^2)\frac{1}{z}.
 \ee
but one also has to take into account the $\varepsilon^4$ term in
$\beta_l/\beta_h$, Eq.\nr{temperature}. The result is:
 \be
  \Delta S^{E}=-{V_3\beta_h \over 16 \pi G_5}
  \bigg(\int_{0}^{z_0}dz (L_h-L_l)-
    \int_{z_0}^{\infty}dz L_l-C\bigg),
  \ee
where $C=-203c^2/20+1/z_0^4$
denotes the sum of the correction terms coming from the boundary term
($-58c^2/5$) and the temperature matching ($+29c^2/20+1/z_0^4$).

\begin{figure}[!htb]
\begin{center}
\includegraphics[width=0.6\textwidth]{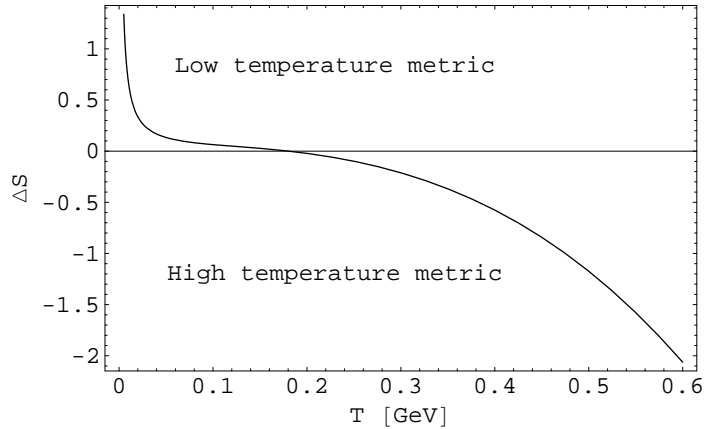}
\end{center}
\vspace{0 cm} \caption{\small The difference $S^E_h - S^E_l$ between
the actions computed with the metrics \nr{hightempmetric} and
\nr{lowtempmetric}. The difference vanishes at $T\approx0.181$ GeV,
above that $S^E_h < S^E_l$ and the high $T$ metric
\nr{hightempmetric} dominates. }\label{HawkingPage}
\end{figure}

Computing numerically the difference of the actions for the metrics
\nr{hightempmetric} and \nr{lowtempmetric} and using the parameter
values obtained earlier from the 4d boundary theory and the function
$w(z)$ in \nr{trialw}, we obtain the result plotted in Fig.
\ref{HawkingPage}, in arbitrary units. The phase transition occurs
when $\Delta S^E=0$ and this corresponds to $T_c\approx 0.181$ GeV.
This is essentially the same as the  $T_c=0.186$ GeV used to
constrain the parameters of the 4d boundary theory. Exact agreement
can be obtained by somewhat modifying the function $w(z)$ in
\nr{trialw}, for example, by writing $w(z)= 2 z^2 e^{-1.041 z^2}-
\tanh z^2$. Again we emphasize that we use the bulk computation to
match the critical temperature obtained from the boundary
computation instead of comparing the full thermodynamic functions to
that of the boundary theory in every detail.

\section{$Q\bar{Q}$ Potential and  Confinement-Deconfinement Transition}

The $Q\bar{Q}$ potential (or free energy at finite $T$) in QCD can
be ``experimentally'' measured using numerical lattice Monte Carlo
techniques. Basically, one measures expectation values of
rectangular Wilson-Polyakov loops with one of the sides becoming
large. At finite $T$, the correlator of two Wilson-Polyakov lines
$\langle\Tr W({\bf L})\Tr W({\bf 0})\rangle =\exp(-F(L,T)/T)$
measure the free energy at finite $T$.

\begin{figure}[!htb]
\begin{center}
\includegraphics[width=\textwidth]{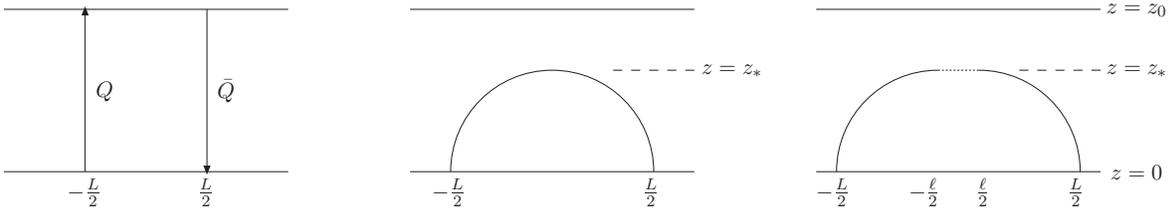}
\end{center}
\vspace{-12cm} \caption{\small Three types of solutions: a $Q\bar Q$
solution (left), a solution with $z'=0$ only at $z=z_*$ (middle), a
solution with a constant piece at $z=z_*$ (right).
}\label{fig:configs}
\end{figure}

To evaluate the $Q\bar{Q}$ potential we use standard gauge
theory/gravity duality techniques (see, e.g., \cite{yee,yeerey}) to
evaluate the finite $T$ expectation value of a Wilson-Polyakov loop
of quarks of fundamental representation. Consider first a
temporal-spatial loop. The Nambu-Goto action of a fundamental string
then is
 \be
  S_{\rm NG}= {1\over 2 \pi \alpha'} \int d\tau d\sigma \sqrt{\det
  G_{MN}\partial_a X^M \partial_b X^N }.
 \ee
Since we are interested in a temporal Wilson loop we choose the
string coordinates as $X^M=(t,x,0,0,z)$, where $z=z(t,x)=z(x)$ is
independent of time. The end points of the string are fixed at
$z(0)=\pm L/2$ and $z(-x)=z(x)$. With the gauge choice of $\tau=t$
and $\sigma= x$ the string action becomes
 \ba
  S_{\rm NG}&=& {\CL^2 \over 2 \pi\alpha'} \Delta t\int_{-L/2}^{L/2} dx {e^{ f(z)} \over z^2}
  \sqrt{ z'(x)^2 + g(z)}\nn
  &=&
  {\CL^2 \over  \pi\alpha'}\Delta t \int_\epsilon^{z_*(L)} dz {e^{ f(z)} \over z^2}
  \sqrt{ 1 + g(z) x'(z)^2},
  \label{sNG}
 \ea
where $z_*$ is the maximum value of $z$ (Fig\ref{fig:configs}) and
$\epsilon$ is a parameter used to regulate the UV divergence at
$z=0$. A potential will be defined by $S_\rmi{NG}=V(L)\Delta t$.

The string frame deformation $f(z)$ is given in terms of the Einstein
frame deformation $h(z)$ by Eq.\nr{fz}, $f(z)=h(z)+4\Phi(z)/3$. Using Eqs.
\nr{hightempmetric}, \nr{lowtempmetric} and \nr{phi} we have
\ba
f(z)&=& (\phi-c)z^2 ={29\over20}[-w(\infty)]cz^2\quad {\rm high}\,T\,{\rm metric},
\label{fhz}\\
&=&{29\over20}[w(\sqrt{c}z)-w(\infty)]cz^2\quad\quad\quad {\rm
low}\,T\,{\rm metric}, \label{flz} \ea where we earlier fixed
$w(\infty)=-1$ and $c=0.127$ GeV$^2$. Due to the rapidly growing
dilaton term the string frame deformation thus behaves qualitatively
differently from the Einstein frame one: it grows monotonically
$\sim +z^2$ for the high $T$ metric and is always $>0$ for the low
$T$ metric.

The equation of motion following from \nr{sNG} is simple and its
first integral is
 \be
  {e^{f(z)} \over z^2} {g(z) \over \sqrt{z'(x)^2 + g(z)}}
  ={\rm constant} = {e^{ f(z_*)} \over z_*^2}\sqrt{g(z_*)},
  \label{eom}
 \ee
where the constant is given as the value of the left hand side when
$z'(x)=0$. From the symmetry of the problem this is at $x=0$, i.e.,
$z_*$ is the maximum value of $z$. Basically the solution can be of
the three types as in Fig. \ref{fig:configs}.

At high temperature with the metric \nr{hightempmetric},  the
equation of motion \nr{eom} has the following types of solutions
(Fig\ref{fig:configs}):
\begin{itemize}
\item A quark-antiquark solution (Fig.\ref{fig:configs} left)
 \be
  x=\pm L/2, \quad x'(z)=0,\quad z'(x)=\infty,\quad z_*=z_0,
 \ee
with the action and the corresponding potential, regulated as
discussed below,
 \ba
  V_{Q\bar Q}&=& {\CL^2
  \over  \pi\alpha'} \int_\epsilon^{z_0} dz {e^{ f(z)} \over z^2}\nn
  &=& {\CL^2 \over  \pi\alpha'}\left[{1\over\epsilon}+
  \int_\epsilon^{z_0} dz {1 \over z^2}\left(e^{
  f(z)}-1\right)-{1\over z_0}\right]\\
  &\Rightarrow&{\CL^2 \over  \pi\alpha'}\left[
  \int_0^{z_0} dz {1 \over z^2}\left(e^{
  f(z)}-1\right)-{1\over z_0}\right].
  \label{sqqbar}
 \ea

\item A solution connecting the quark and antiquark with $z'(x)=0$ only at $x=0$
(Fig.\ref{fig:configs} middle). Integration of \nr{eom} relates
$z_*$ and $L$:
 \be
  L=2 \int_0^{z_*} dz {1 \over \sqrt{g(z) \left[q(z_*)/q(z)-1\right]}},
  \label{L}
 \ee
where
 \be
   q(z)=e^{-2f(z)}{z^4\over z_0^4-z^4},
  \label{gzh}
 \ee
for high temperature phase and
 \be
   q(z)=e^{-2f(z)}z^4,
  \label{gzl}
 \ee
for low temperature phase. The corresponding potential is,
regulating the $z\to0$ divergence as discussed below,
 \ba
  V(L,z_0;f)&=& { \CL^2 \over \pi\alpha'}
  \int_\epsilon^{z_*} {dz\over z^2}\,  e^{f(z)}
 {1\over\sqrt{1-q(z)/q(z_*)}}\nn
 &=&{ \CL^2 \over \pi\alpha'}
 \left[{1\over\epsilon}+\int_\epsilon^{z_*} {dz\over z^2}
 \left(e^{f(z)}{ 1\over\sqrt{1-q(z)/q(z_*)}} -1\right)-{1\over z_*}\right]\\
 &\Rightarrow&{ \CL^2 \over \pi\alpha'}
 \left[\int_0^{z_*} {dz\over z^2}
 \left(e^{f(z)}{ 1\over\sqrt{1-q(z)/q(z_*)}} -1\right)-{1\over z_*}\right]
 \label{V}
 \ea

\item One solution of the equation of motion also is (Fig.\ref{fig:configs} right)
 \be
   z=z_*={\rm constant}.
 \ee
This can be made a solution of required type by connecting its end
points to the points $x(0)=\pm L/2$ by a solution of the previous
type with nonzero $z'(x)$. If the length of the constant piece is
$\ell$, the value of $z_*$ will then be determined by \nr{L} with
$L\rightarrow L-\ell$ and the total potential is \nr{V} evaluated
for this $z_*(L-\ell)$ added to
 \be
 \label{linear}
   V_\rmi{planar}={ \CL^2 \over
   2\pi\alpha'}{e^{f(z_*)}\over z_*^2}\sqrt{g(z_*)}\,\,\ell.
 \ee
For fixed $L=\ell + (L-\ell)$, $0\le \ell<L$, we thus obtain a
family of solutions with increasing length $\ell$ of the constant
region, decreasing lengths $(L-\ell)/2$ of the two connecting
regions at the ends and decreasing height $z_*$. It appears that at
fixed $L$ all these solutions have a larger $V$ than the one of the
middle type.

\end{itemize}

\begin{figure}[!htb]
\begin{center}
\includegraphics[width=0.45\textwidth]{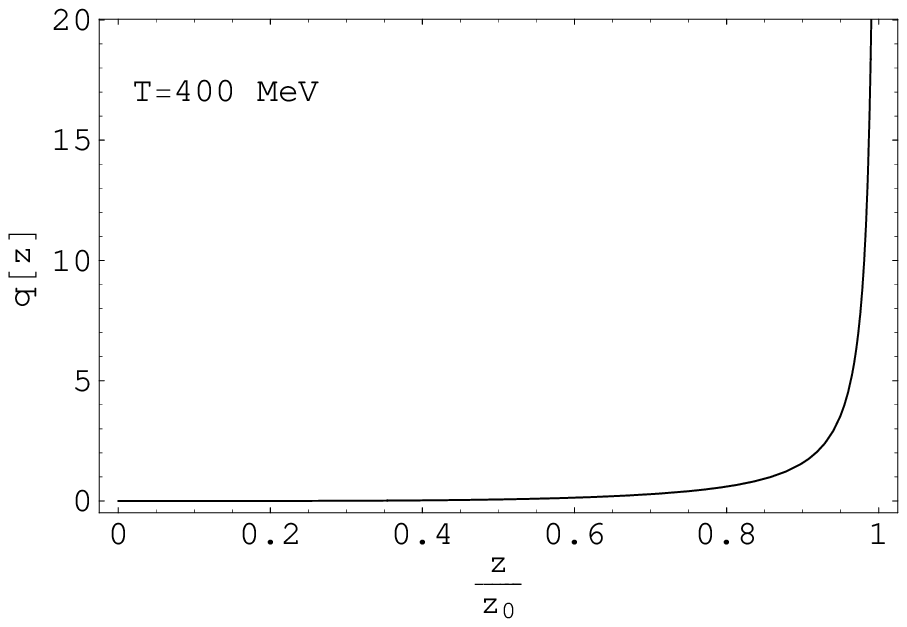} \hspace{1 cm}
\includegraphics[width=0.45\textwidth]{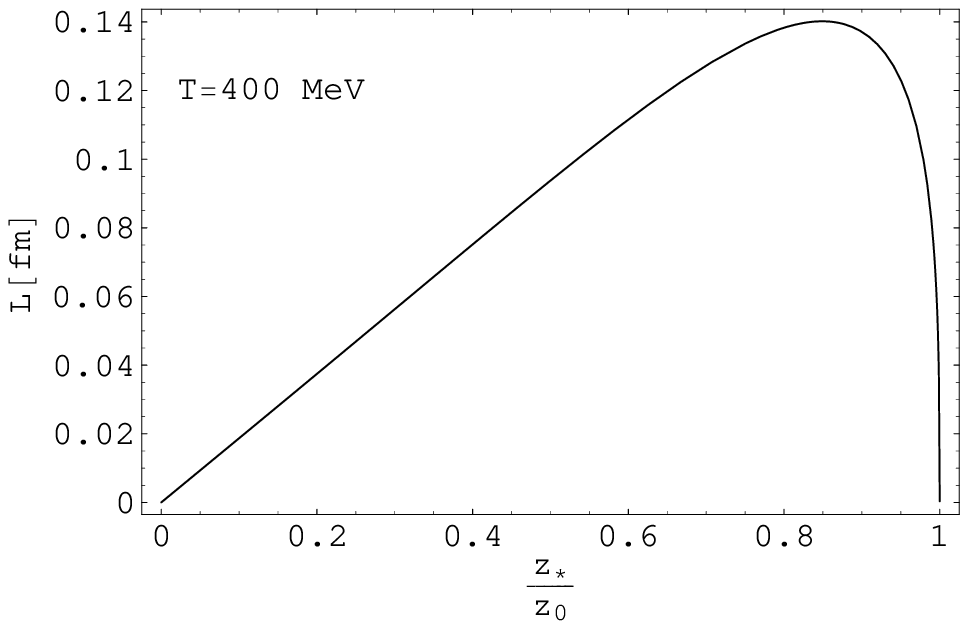} \\
\includegraphics[width=0.45\textwidth]{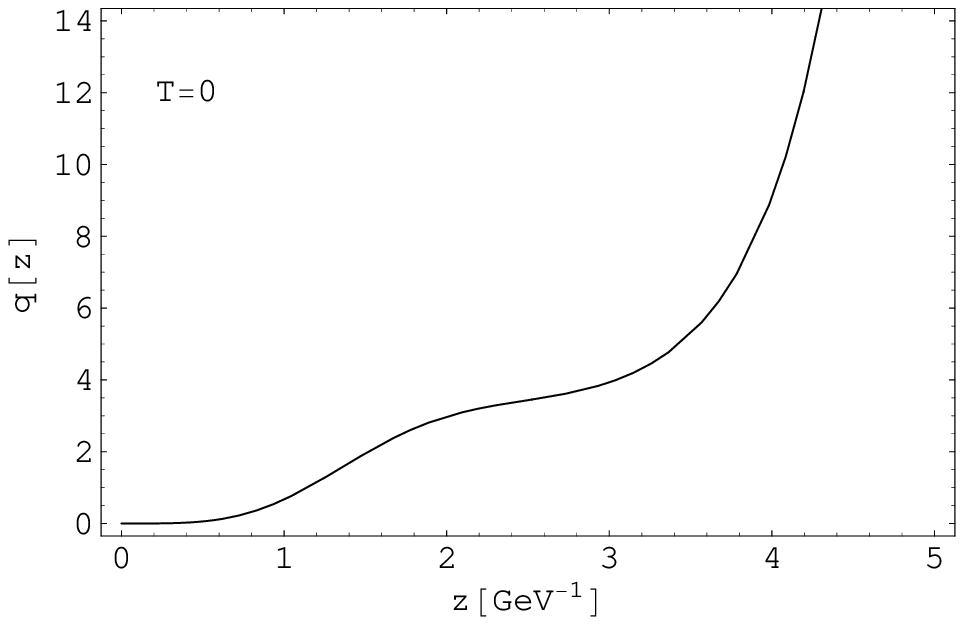} \hspace{1 cm}
\includegraphics[width=0.45\textwidth]{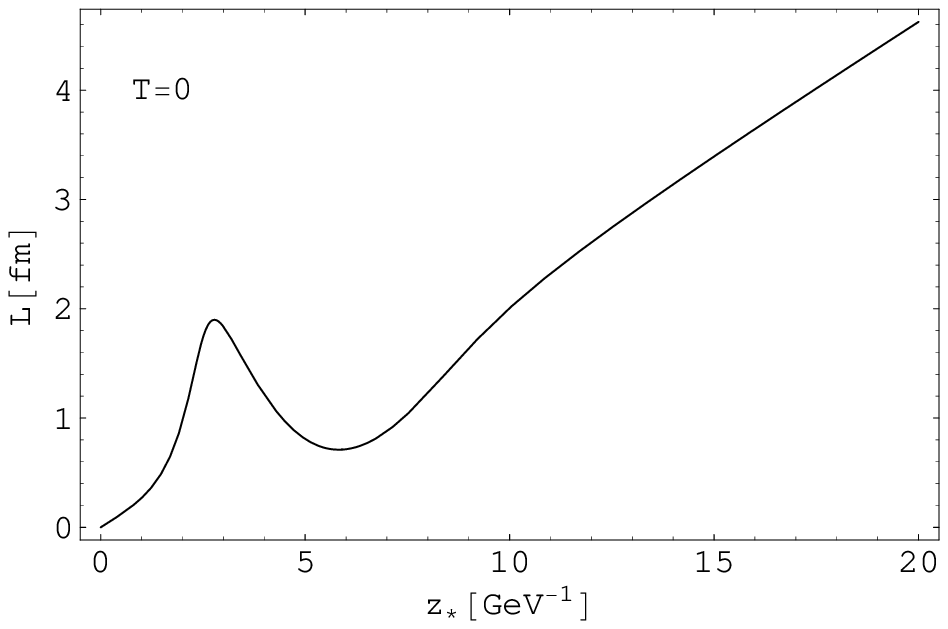}
\end{center}
\caption{\small  The functions $q(z)$ in Eq.\nr{gzh} and $L(z_*)$ in
Eq.\nr{L} for the high temperature metric with  $T=400 {\rm MeV}$
using the deformation \nr{fhz} (upper two panels). The lower two
panels show the same for the low temperature metric with $T=0$, now
plotted vs $z$ in units of 1/GeV.}\label{Lfigs}.
\end{figure}

Eqs. \nr{sqqbar} and \nr{V} diverge for $z\to0$ and have to be
renormalized. There are two possibilities:
\begin{itemize}

\item[{1.}] Neglect the $1/\epsilon$ terms in \nr{sqqbar} and \nr{V} and put $\epsilon=0$ in
the remainder;

\item[{2.}] Subtract the two and put $\epsilon=0$ in the remainder, i.e.,
redefine $V$ as $V-V_{Q\bar Q}$.

\end{itemize}

We use the first one and the final potential is given by combining
Eqs. \nr{L}, \nr{sqqbar}
and \nr{V} so that the equilibrium state always corresponds to the
smaller one of $V$ and $V_{Q\bar Q}$.

\begin{figure}[!htb]
\begin{center}
\includegraphics[width=0.49\textwidth]{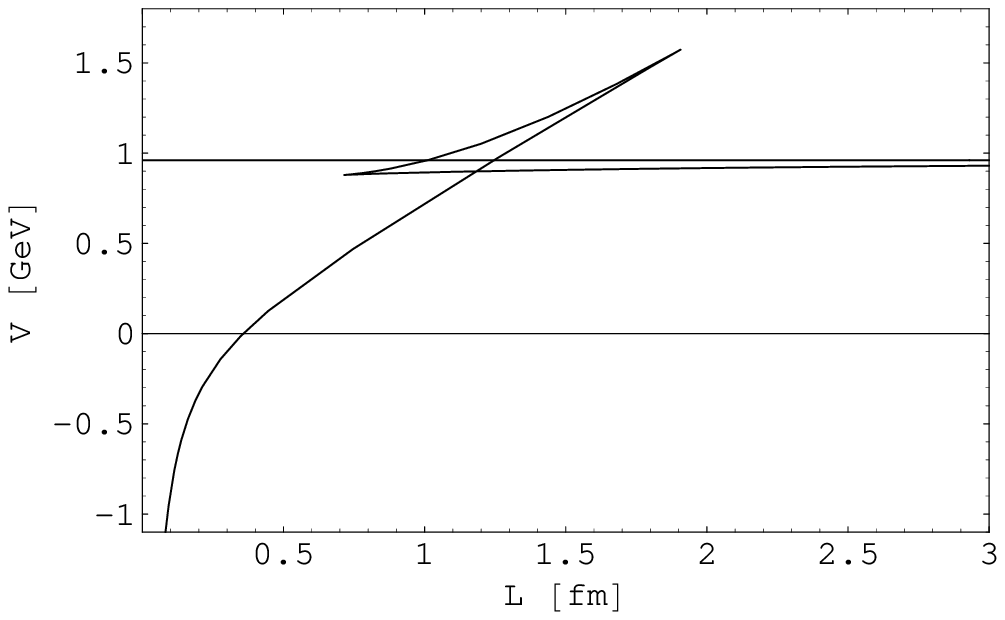}
\includegraphics[width=0.49\textwidth]{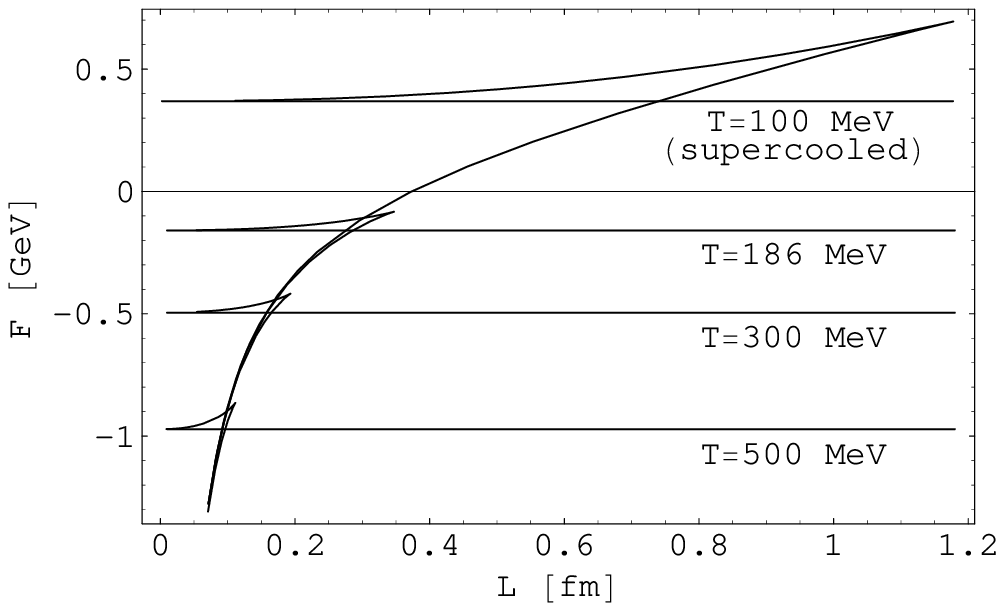}
\end{center}
\caption{\small $V(L)$ vs $L$ as computed from eq. \nr{V} for the
low temperature metric (left, $T$-independent) and $F(L,T)$ vs
$L$for the high temperature metric (right) at
$T=500,300,T_c=186,100$ MeV. Note the difference in $L$-scales. The
almost horizontal part of the low $T$ potential comes from the
$z_*>6  $ part of the $L(z_*)$-curve in Fig.\ref{Lfigs}, the
horizontal parts of the high $T$ potential are $L$-independent
values of $V_{Q\bar Q}$ from \nr{sqqbar}. }\label{fig:VL}
\end{figure}

Consider first the low $T$ metric, using the deformation \nr{flz}
and parameters obtained earlier. We plot $q(z)$ and $L(z_*)$ with
$z$ and $z_*$ in units of 1/GeV in Fig. \ref{Lfigs}. The structure
in $L$ at small $z_*$ is significant and leads to up to three values
of $V$ corresponding to one $L$ as seen in the plot of the
($T$-independent) potential $V(L)$ vs $L$ in Fig. \ref{fig:VL}.
Tracing through the curve $L=L(z_*)$ in Fig. \ref{Lfigs} one can
trace through the curve $V=V(L)$ in Fig. \ref{fig:VL}. The almost
constant branch of $V(L)$ arises from the $z_*>6 {\rm GeV}^{-1}$
part of $L(z_*)$.

We interpret the resulting $T=0$ potential as showing a confining part for $L<1.2$ fm
and a constant part for $L>1.2$ fm.

For $L < 1.2 \,\,{\rm fm}$
we can express the potential in Cornell form
\be
  V(L) = -{4 \over 3} {\alpha_s \over L}+\sigma L.
 \ee
If we use $\alpha_s \approx 0.36$, then we get the AdS
radius $ \CL^2/ (\pi\alpha') \approx 0.67$ and $\sigma \approx
0.168{\rm GeV}^2$ or $\sqrt{\sigma} \approx 410 {\rm MeV}$. This is
in reasonable agreement with
recent lattice data giving $\sqrt{\sigma} \approx 460 $ MeV\cite{karsch}.
Note that $T_c/\sqrt{\sigma} \approx 0.46$ in this case.

Beyond 1.2 fm the potential is approximately constant and grows
slowly from $\approx$ 0.9 GeV to $V_{Q\bar Q}=0.961$ GeV, obtained
by evaluating Eq.\nr{sqqbar} with $z_0=\infty$. One can speculate
that the constant potential is an indication of QCD string breaking
of dynamical quarks\cite{kkw}. The constant part starts at 1.2 fm
and its energy scale is about $ 1$ GeV. This should be in principle
the twice the mass of the lightest hadron. Current lattice data
indicates that the string breaking occurs about $ 1.2-1.4 {\rm fm}$
at the energy scale of $1.0-1.2$ GeV \cite{kaczmarek}. This is
consistent with our model calculation. The reason why we see the
string breaking is not obvious at first thought because it happens
only when dynamical quarks are included. However, from the
consideration of linear Regge behavior of meson mass spectrum and
the boundary thermodynamics we have been lead to a specific form of
the function $w(z)$ containing a scale related to the mass scale of
the fundamental matter in QCD. Thus in our model (\ref{bulkmetric})
we are already implicitly including dynamical quarks from the
beginning. In AdS/CFT, light dynamical quarks  can be added by
introducing a spacetime-filling D7 brane \cite{karchkatz}. It would
be very interesting to see in supergravity whether the original
metric ans\"atz \nr{bulkmetric} can be related to this flavor brane
background.

Now that we have determined the AdS radius
$ \CL^2/ (\pi\alpha') \approx 0.67$,
we can draw the $Q\bar{Q}$
potential at high temperatures, again using the deformation \nr{fhz}
and parameters determined earlier.
The quantities $q(z)$ and $L(z_*)$ with $z$ and $z_*$ in units of $z_0$
are plotted in Fig. \ref{Lfigs} and $V(L)$ in Fig. \ref{fig:VL} for
$T=400$ MeV. It is again useful to trace through $L(z_*)$ and see how it
maps to $V(L)$: one starts from $L=0,\,V=-\infty$, then $L$ and $V$
grow to a maximum value after which both again decrease but so that
$V$ is almost constant. Finally again $L=0$ at $z_*=z_0$ at which
point $V(L=0)=V_{Q\bar Q}$. This is $L$-independent and is shown as
the horizontal curves in Fig.\ref{fig:VL}.

One can observe the following for the high temperature phase:
\begin{itemize}
\item[{1.}] The overall potential is the lowest of the curves/line and contains the usual
conformally invariant $V\sim -1/L$ at small $L$. At some $L$ this
reaches $V_{Q\bar Q}$ and from this $L$ onwards the preferred state
is the $Q\bar Q$ state with a constant $L$-independent potential.
This is the structure observed already in \cite{yee} for an
undeformed metric.
\item[{2.}] At very large $T$, small $z_0$, the large $L$ value of the potential
(free energy) $V_{Q\bar Q}$ from Eq.\nr{sqqbar} approaches $\CL^2/(
\pi\alpha')(-\pi T)$. Although one cannot strictly separate internal
energy $E$ and entropy $S$ in the free energy of the $Q\bar Q$
state, this is compatible with $E=0, S= \CL^2/( \pi\alpha')\cdot \pi
\approx 2.1$. Notice that $\Delta S \sim {\cal O}(1)$ for simple
quark and antiquark system.
\item[{3.}] As the temperature decreases below $T_c$, a first order phase
transition occurs as we have seen in sections 3 and 5. One can continue
plotting $V(L)$ in this metastable phase, as done in Fig.\ref{fig:VL}, and obtain a linear confining
potential. However, the stable phase is the one corresponding to the
metric \nr{lowtempmetric} which should be used for $T<T_c$.

\end{itemize}

\section{Discussion}

In this paper, we have considered a 5d gravity action which
presumably could descend from a higher dimensional fundamental
theory. Following \cite{karch}, we proposed a solution of the
equations of motion that is dual to a QCD-like gauge theory on the
4d boundary. Using standard gravity/gauge theory duality
correspondence we computed the 4d boundary energy-momentum tensor.
The gauge dynamics on the boundary is that of a strongly coupled
fluid modified by a vacuum energy. The bulk geometry was further
constrained so that the 4d boundary theory shows a first order phase
transition. With constraints from the hadron spectrum, the
parameters of the deformation were fixed. Then we considered the 5d
bulk transition following the recipe of Hawking and Page
\cite{hawkingpage} and found that the geometric transition in the 5d
bulk precisely matched with the phase transition on the 4d boundary.
This is a remarkable correspondence and shows the power of
gravity/gauge theory duality and the reasonableness of the model we
have set up for the dual construction of QCD-like theories. The
computation of quark-antiquark free energy (potential at $T=0$)
showed that at $T=0$ potential contains a confining linear part but
also string breaking, indicating that the deformation contains
effects of dynamical quarks. At $T>T_c$ we saw a deconfined state
with entropy = 2.1/pair.

In general terms, the whole setup started from three unknown functions,
the deformations of the high $T$ and low $T$ metrics and the dilaton.
We chose simple ans\"atze for the high $T$ deformation, $h(z)=-cz^2$,
and for the dilaton, $\Phi(z)=\phi z^2$. The low $T$ deformation was
then constrained in various ways and one ended up with the
function $w(z)$ plotted in Fig.\ref{wz}. There is some freedom to
change this function in the intermediate $z$ range, but not much
without destroying the coincidence of the 4d boundary and 5d bulk
transitions.

Many QCD-like features treated in this paper are basically IR
dominant phenomena, but since from gravity/gauge theory duality, the
boundary data is critically dependent on the boundary behavior of
the bulk metric, the metric form in the UV region near the boundary
becomes as important. If we solve the equations of motion of
supergravity in top-down approach, this UV/IR behavior is naturally
encoded in the solution, but when we approach phenomenologically we
have to put this UV/IR behavior by hand. What we saw here is that
the form of the metric is stringently constrained by this.

There are several directions of future study. First is to check the
our ans\"atz  using various physical quantities in QCD and refine
the
 form of $w(z)$. Also it would be highly desirable to reproduce the
solution in supergravity. Even though this is very difficult task,
we have some better guiding principle at least.

\vskip6mm
\noindent

\subsection*{Acknowledgement}

We thank Esko Keski-Vakkuri, Mikko Laine and Kari Rummukainen for
helpful discussions and Kostas Skenderis for valuable comments. This
research has been supported by Academy of Finland, contract number
109720.


\end{document}